\documentclass[aps,prd,superscriptaddress,showpacs,preprintnumbers]{revtex4}
\usepackage{graphicx,color,here}

\newcommand{\be}{\begin{equation}}
\newcommand{\ee}{\end{equation}}
\newcommand{\bea}{\begin{eqnarray}}
\newcommand{\eea}{\end{eqnarray}}

\newcommand{\bfk}{\mbox{\boldmath $k$}}
\newcommand{\bfq}{\mbox{\boldmath $q$}}
\newcommand{\bfP}{\mbox{\boldmath $P$}}

\newcommand{\bfp}{\mbox{\boldmath $p$}}

\def\lsim{\mathrel{\rlap{\lower4pt\hbox{\hskip1pt$\sim$}}\raise1pt\hbox{$<$}}}
\def\gsim{\mathrel{\rlap{\lower4pt\hbox{\hskip1pt$\sim$}}\raise1pt\hbox{$>$}}}
\def\nostrocostruttino#1\over#2{\mathrel{\mathop{\kern 0pt \rlap
{\hbox{$#1$}}} \hbox{\kern-.135em $#2$}}}

\newcommand{\NP}[1]{{\it Nucl.\ Phys.}\ {\bf #1}}
\newcommand{\ZP}[1]{{\it Z.\ Phys.}\ {\bf #1}}
\newcommand{\PL}[1]{{\it Phys.\ Lett.}\ {\bf #1}}
\newcommand{\PR}[1]{{\it Phys.\ Rev.}\ {\bf #1}}
\newcommand{\PRL}[1]{{\it Phys.\ Rev.\ Lett.}\ {\bf #1}}

\def\kt{k_\perp}

\def\ptq{p_\perp}
\def\pt{P_T}
\def\xb{x_{_{\!B\!j}}}
\def\xp{x^\prime}
\def\zp{z^\prime}
\begin{document}
\title{Semi-Inclusive Deep Inelastic Scattering processes from small
to large $\bfP_T$}

\author{M.~Anselmino}
\affiliation{Dipartimento di Fisica Teorica, Universit\`a di Torino and \\
          INFN, Sezione di Torino, Via P. Giuria 1, I-10125 Torino, Italy}
\author{M.~Boglione}
\affiliation{Dipartimento di Fisica Teorica, Universit\`a di Torino and \\
          INFN, Sezione di Torino, Via P. Giuria 1, I-10125 Torino, Italy}
\author{A.~Prokudin}
\affiliation{Dipartimento di Fisica Teorica, Universit\`a di Torino and \\
          INFN, Sezione di Torino, Via P. Giuria 1, I-10125 Torino, Italy}
\author{C.~T\"{u}rk}
\affiliation{Dipartimento di Fisica Teorica, Universit\`a di Torino and \\
          INFN, Sezione di Torino, Via P. Giuria 1, I-10125 Torino, Italy}


\begin{abstract}
We consider the azimuthal and $P_T$ dependence of hadrons produced in
unpolarized Semi-Inclusive Deep Inelastic Scattering (SIDIS) processes, within
the factorized QCD parton model. It is shown that at small $P_T$ values, $P_T
\lsim 1$ GeV/$c$, lowest order contributions, coupled to unintegrated
(Transverse Momentum Dependent) quark distribution and fragmentation functions,
describe all data. At larger $P_T$ values, $P_T \gsim 1$ GeV/$c$, the usual
pQCD higher order collinear contributions dominate. Having explained the full
$P_T$ range of available data, we give new detailed predictions concerning the
azimuthal and $P_T$ dependence of hadrons which could be measured in ongoing or
planned experiments by HERMES, COMPASS and JLab collaborations. \noindent

\end{abstract}


\maketitle

\section{\label{Intro} Introduction}

In Ref.~\cite{sidis1} a comprehensive analysis of Semi-Inclusive Deep
Inelastic Scattering (SIDIS) processes within a factorized QCD parton model
at $\mathcal{O}(\alpha _s ^0)$ was performed in a kinematical scheme in
which the intrinsic tranverse momenta of the quarks inside the initial
proton ($\bfk_\perp$) and of the final detected hadron with respect to the
fragmenting quark ($\bfp_\perp$) were fully taken into account. The dependence
of the unpolarized cross section on the azimuthal angle $\phi_h$ between the
leptonic and the hadron production plane (Cahn effect \cite{cahn}) was
compared to the available experimental data, and used to estimate the average
values of $\langle k_\perp^2 \rangle$ and $\langle p_\perp^2 \rangle$. These
values were adopted in modeling the intrinsic motion dependence of the
quark distribution and fragmentation functions. This allowed a consistent
description of the azimuthal dependence observed by HERMES and COMPASS
collaborations in SIDIS off transversely polarized protons \cite{herm,comp},
with the subsequent extraction \cite{sidis1,sidis2} of the Sivers distribution
function \cite{siv}.

In Ref. \cite{sidis1} the main emphasis, following the original idea of Cahn,
was on the role of the parton intrinsic motion, with the use of unintegrated
quark distribution and fragmentation functions. That applies to large $Q^2$
processes, in a kinematical regime in which $P_T \simeq \Lambda_{\rm QCD}
\simeq \kt$, where $P_T = |\bfP_T|$ is the magnitude of the final hadron
transverse momentum. In this region QCD factorization with unintegrated
distributions holds \cite{ji} and lowest order QED elementary processes, $\ell
\, q \to \ell \, q$, are dominating: the soft $P_T$ of the detected hadron is
mainly originating from quark intrinsic motion \cite{EMC1,EMC2,E665}, rather
than from higher order pQCD interactions, which, instead, would dominantly
produce large $P_T$ hadrons \cite{gp,chay,maniatis,sassot}.

Indeed, a look at the results of Ref. \cite{sidis1} (see, in particular, Figs.
5 and 6) immediately shows that, while the inclusion of intrinsic $k_\perp$ and
$p_\perp$ -- coupled to lowest order partonic interactions -- leads to an
excellent agreement with the data for small values of the transverse momentum
$P_T$ of the final hadron, it badly fails at higher $P_T$: the turning point is
around $P_T \sim 1$ GeV/$c$. A similar conclusion was drawn in Ref.
\cite{chay}. The large $P_T$ region has been discussed at length in the
literature and is related to contributions from higher order QCD processes,
like hard gluonic radiation and elementary scatterings initiated by gluons:
these cannot be neglected when $P_T \gg \lambda _{\rm QCD}$
\cite{gp,chay,maniatis,sassot}.

In this paper we start by showing that a complete agreement with data in the
full range of $P_T$ can be achieved; for $P_T \lsim 1$ GeV/$c$ we follow the
approach of Ref. \cite{sidis1} -- $P_T$ originated by the intrinsic $k_\perp$
and $p_\perp$ with $\mathcal{O}(\alpha _s ^0)$ partonic interaction -- while in
the range of $P_T \gsim 1$ GeV/$c$ we add the pQCD contributions -- collinear
partonic configurations with higher order [up to $\mathcal{O}(\alpha_s^2)$]
partonic interactions which generate the large $P_T$. We shall see that indeed
most available data can be explained; the intrinsic $k_\perp$ contributions
work well at small $P_T \lsim 1$ GeV/$c$ and fail above that, while the higher
order pQCD collinear contributions explain well the large $P_T \gsim 1$ GeV/$c$
data and fail, or are not even applicable, below that. The two contributions
match in the overlapping region, $P_T \sim 1$ GeV/$c$, where it might be
difficult to disentangle one from the other, as they describe the same physics.
Infact parton intrinsic motions originate not only from confinement, but also
from soft gluon emission, which, due to QCD helicity conservation, cannot be
strictly collinear. Similar studies, concerning single transverse-spin
asymmetries in Drell-Yan and SIDIS processes, with separate contributions --
TMD quark distributions and higher-twist quark gluon correlations -- in
separate regions, have recently been published \cite{werner1,werner2}.

Having achieved such a complete understanding of the $P_T$ dependence of
the SIDIS cross sections we obtain a full confidence on the regions of
applicability of the two approaches. We re-analyse the azimuthal
dependence of the unpolarized cross section -- the Cahn effect, described
in Ref. \cite{sidis1} -- which depends on quantities integrated over $P_T$.
The actual data are dominated by the low $P_T$ contributions, and the results
previously obtained remain valid; we obtain slightly different values
of the parameters $\langle k_\perp^2 \rangle$ and $\langle p_\perp^2 \rangle$.
We then consider running experiments (HERMES, COMPASS and experiments at
JLab) and physical observables which are being or will soon be measured.
They are mainly in the small $P_T$ regions and we give full sets of
predictions for them.

The plan of the paper is the following: in Section \ref{model} we give a short
overview of the kinematics and a collection of the basic formulae needed for
the computation of the SIDIS cross sections, both in the low $P_T$ approach of
Ref. \cite{sidis1} and in the pQCD large $P_T$ region; in Section~\ref{results}
we discuss and compare our results for $d\sigma/dP_T$, $d\sigma/d\phi_h$ and
$\langle \cos\phi_h \rangle$ with the existing experimental data, over a very
wide range of $P_T$ values; in Section~\ref{predictions} we give predictions
for the forthcoming measurements of $d\sigma/dP_T$ and $\langle \cos\phi_h
\rangle$ at HERMES, Compass and JLab. Some considerations on $\langle
\cos2\phi_h \rangle$ are made. Finally, in Section~\ref{conclusions} we draw
our conclusions.

\section{\label{model} Kinematics and Cross Sections}

\begin{figure}[t]
\begin{center}
\scalebox{0.35}{\input{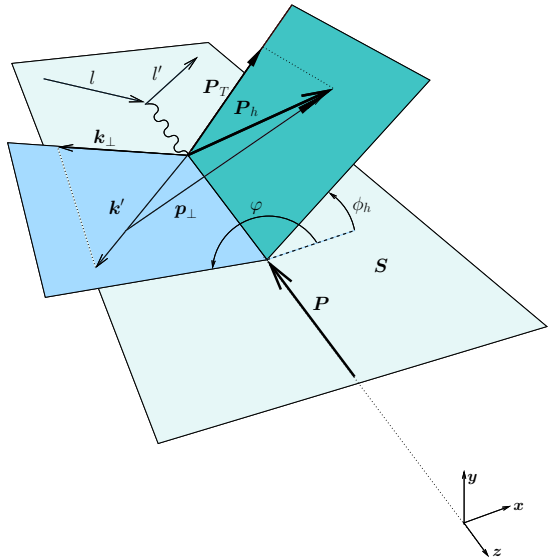}}
\caption{\label{fig:planessidis} Three dimensional kinematics of the SIDIS
process.}
\end{center}
\end{figure}

We consider SIDIS processes $\ell \, p \to \ell \, h \, X$ in the $\gamma^* p$
c.m. frame, as shown in Fig. \ref{fig:planessidis}. The photon and the proton
collide along the $z$ axis with momenta $\bfq$ and $\bfP$ respectively; the
leptonic plane coincides with the $x$-$z$ plane. We adopt the usual SIDIS
variables (neglecting all masses):
\bea s = (P + \ell)^2 \quad\quad (P+q)^2 = W^2 = \frac{1-\xb}{\xb}\,Q^2
\quad\quad  q^2 = -Q^2 \nonumber \\
\xb = \frac {Q^2}{2P \cdot q} = \frac{Q^2}{W^2 + Q^2} \quad\quad y = \frac{P
\cdot q}{P \cdot \ell} = \frac{Q^2}{\xb s} \quad\quad z_h = \frac {P \cdot
P_h}{P \cdot q} \> \cdot \label{kin} \eea
%
%

The SIDIS differential cross section can schematically be written in terms of
a perturbative expansion in orders of $\alpha_s$ as follows
\be
d\sigma \;=\; \alpha_s ^0 \;d\sigma_0 +  \
          \alpha_s ^1 \;d\sigma_1 +
          \alpha_s ^2 \;d\sigma_2 + ...\,,
\label{pert-exp} \ee
where $d\sigma$ is a short hand notation to indicate
\(\displaystyle{\frac{d^5\sigma^{\ell p \to \ell h X }}{d\xb \, dy \, dz_h \,
d^2 \bfP _T}} = \xb s \, {\frac{d^5\sigma^{\ell p \to \ell h X }}{d\xb \, dQ^2
\, dz_h \, d^2 \bfP _T}} \cdot\)

The first term in this expansion is the lowest order one; in the elementary
interaction, $\ell \, q \to \ell \, q$, a virtual photon with four-momentum $q$
strikes a quark which carries a transverse momentum
$\bfk_\perp=k_\perp(\cos\varphi,\sin\varphi,0)$ in addition to a fraction $x$
of the light-cone proton momentum. The final detected hadron $h$ originates
from the fragmentation of the outgoing quark: $\bfp _\perp$ is the transverse
momentum of $h$ {\it with respect to the direction} of the fragmenting quark
and $z$ is the fraction of the light-cone quark momentum carried by the
resulting hadron. Consequently, the detected hadron can have a transverse
momentum $\bfP_T$ with a magnitude $P_T \simeq \langle \kt \rangle \simeq
\langle p_\perp \rangle$. Indeed, this is the main source of hadrons with a
small value of $P_T$ \cite{chay,EMC1,EMC2,E665}.

Such a mechanism translates into a factorized \cite{ji} expression for the
SIDIS cross section, valid at all orders in $(k_\perp/Q)$:
\bea \frac{d^5\sigma^{\ell p \to \ell h X }_0} {d\xb \, dQ^2 \, dz_h \, d^2
\bfP_T} && = \sum_q \int {d^2 \bfk _\perp}\; f_q(x,\bfk _\perp) \; \frac{d
\hat\sigma ^{\ell q\to \ell q}}{dQ^2} \; J\; \frac{z}{z_h} \; D_q^h(z,\bfp
_\perp)
\nonumber \\
&& = \sum_q  e_q^2 \int  \! d^2 \bfk _\perp \; f_q(x,\bfk _\perp) \;
\frac{2\pi\alpha ^2}{\xb^2 s^2}\,\frac{\hat s^2+\hat u^2}{Q^4}\; D_q^h(z,\bfp
_\perp) \; \frac{z}{z_h} \, \frac{\xb}{x}\left( 1 +
\frac{\xb}{x}\frac{\kt^2}{Q^2} \right)^{\!\!-1} \> , \label{sidis-Xsec-final}
\eea
as explained in Ref.~\cite{sidis1}, where the exact relationships between $x,
z, \bfp_\perp$ and the observables $\xb, z_h, \bfP_T$ are given. Notice that at
$\mathcal{O}(\kt/Q)$ one has $\xb = x$, $z_h = z$ and $\bfP_T = z\bfk_\perp +
\bfp_\perp$. $f_q(x,\bfk _\perp)$ and $D_q^h(z,\bfp _\perp)$ are the parton
density and the fragmentation function respectively, for which we assume the
usual $x, k_\perp$ or $z, p_\perp$ factorization, with a gaussian $k_\perp$ and
$p_\perp$ dependence:
\be f_q(x,\bfk_\perp) = f_q(x) \, \frac{1}{\pi \langle\kt^2\rangle} \,
e^{-{\kt^2}/{\langle\kt^2\rangle}} \quad\quad D_q^h(z,\bfp _\perp) = D_q^h(z)
\, \frac{1}{\pi \langle p_\perp^2\rangle} \, e^{-{p_\perp^2}/\langle
p_\perp^2\rangle}\,,\label{partondf} \ee
so that
\be
\int d^2\bfk_\perp \> f_q(x,\bfk_\perp) = f_q(x) \quad\quad
\int d^2\bfp_\perp \> D_q^h(z,\bfp_\perp) = D_q^h(z) \>.
\ee

The integration over $d^2\bfk_\perp$ in Eq. (\ref{sidis-Xsec-final}) induces a
dependence on $\cos\phi_h$ [at $\mathcal{O}(P_T/Q)$] and on $\cos2\phi_h$ [at
$\mathcal{O}(P_T/Q)^2$], where $\phi_h$ is the azimuthal angle of $\bfP_T$. The
explicit expression, at $\mathcal{O}(P_T/Q)$, is given by:
\bea \nonumber && \frac{d^5\sigma^{\ell p \to \ell h X }}{d\xb \, dQ^2 \, dz_h
\, d^2\bfP _T} \simeq \sum_q \frac{2\pi\alpha^2e_q^2}{Q^4} \> f_q(\xb) \>
D_q^h(z_h) \, \biggl[
1+(1-y)^2 \\
&& \hskip 36pt - 4 \> \frac{(2-y)\sqrt{1-y}\> \langle\kt^2\rangle \, z_h \,
P_T} {\langle\pt^2\rangle \, Q}\> \cos \phi_h \biggr]
\frac{1}{\pi\langle\pt^2\rangle} \, e^{-P_T^2/\langle\pt^2\rangle} \, .
\label{cahn-anal-app} \eea
where, $\langle P_T^2 \rangle = \langle p_\perp^2\rangle +
z^2\langle\kt^2\rangle$.

\begin{figure}
\begin{center}
\includegraphics[width=0.65\textwidth]
{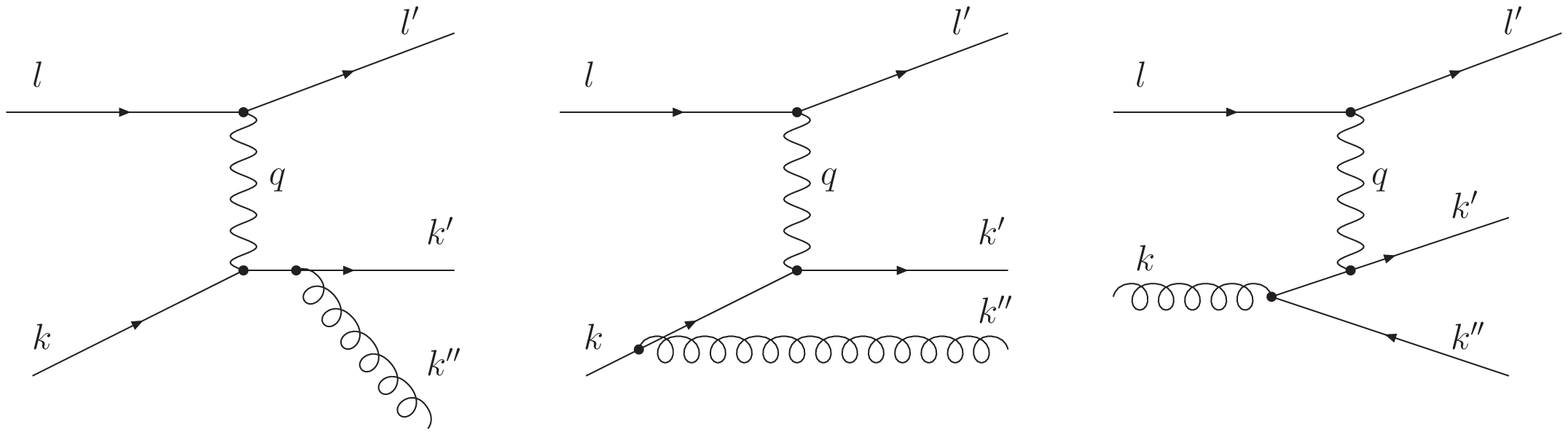}
\end{center}
\caption{\label{feynman} }Feynman diagrams corresponding to $\ell q$
and $\ell g$ elementary scattering at first order in $\alpha_s$.
\end{figure}

Let us now consider the contributions of order $\alpha_s$, $d\sigma_1$
in Eq. (\ref{pert-exp}). We follow the approach of Ref. \cite{chay}. The
relevant partonic processes, shown in Fig.~\ref{feynman}, are those in
which the quark emits a hard gluon or those initiated by gluons:
\be
\gamma^* \,+\, q \to q \,+\, g \quad\quad
\gamma^* \,+\, q \to g \,+\, q \quad\quad
\gamma^* \,+\, g \to q \,+\, \bar{q}\,.
\ee
It is clear that now, contrary to the lowest order QED process,
$\gamma^* \,+\, q \to q$, the final parton can have a large transverse
momentum, even starting from a collinear configuration. Such a contribution
certainly dominates the production of hadrons with large $P_T$ values.

One introduces the parton variables $\xp$ and $\zp$, defined
similarly to the hadronic variables $\xb$ and $z_h$,
\be
\xp = \frac {Q^2}{2k \cdot q}  = \frac{\xb}{\xi} \quad\quad
\zp = \frac {k \cdot k^\prime}{k \cdot q}  = \frac{z_h}{\zeta} \,,
\ee
where $k$ and $k^\prime$ are the four-momenta of the incident and fragmenting
partons respectively. $\xi$ and $\zeta$ are the usual light-cone momentum
fractions, which, in the collinear configuration with massless partons are
given by $k = \xi P$ and $P_h = \zeta k^\prime$. We denote by $\bfp_T$
(not to be confused with $\bfp_\perp$) the transverse momentum, with
respect to the $\gamma^*$ direction, of the final fragmenting parton,
$\bfP_T = \zeta \bfp_T$.

The semi-inclusive DIS cross section, in the QCD parton model with
collinear configuration, can be written, in general, as:
\bea
\label{eq:SIDIScrossoriginal}
\frac{d^5\sigma^{lp\rightarrow lhX}}{d\xb dy \,dz_h \,d^2\bfP_T} &=&
\sum_{i,j} \int d\xp \, d\zp \, d^2\bfp_T \, d\xi \, d\zeta
\delta\left( \xb - \xi \xp\right) \delta\left( z_h - \zeta \zp \right)
\delta^2\left( \mathbf{P}_T - \zeta \bfp_T \right)
\nonumber \\
&& \times
f_i\left(\xi,Q^2\right)
\frac{d \hat{\sigma}_{ij}}{d\xp \,d y \,d\zp \,d^2\bfp_T}
D_j^h\left(\zeta,Q^2\right)\,.
\eea
%
%
To first order in $\alpha_s$ the partonic cross section is given by
\cite{mendez-kroll,chay}
\be
\frac{d \hat{\sigma}_{ij}}{d\xp \,dy \,d\zp \,d^2\bfp_T} =
\frac{\alpha^2e_q^2}{16\pi^2 Q^4} \,y \, L_{\mu\nu} M_{ij}^{\mu\nu}
\delta\left(p^2_T - \frac{\zp}{\xp}(1-\xp)(1-\zp)Q^2\right),
\ee
where $ij$ denote the initial and fragmenting partons, $ij = qq, qg, gq$.
Inserting the above expression into Eq. (\ref{eq:SIDIScrossoriginal}) yields,
for the $\mathcal{O}(\alpha_s)$ cross section:
\be
\frac{d^5\sigma^{lp\rightarrow lhX}_1}{d\xb dy \,dz_h \,d^2\bfP_T} =
\frac{\alpha^2\,e_q^2}{16\pi^2}\, \frac{y}{Q^4}
\int_{\xb}^1 \frac{d\xp}{\xp P_T^2 + z_h^2(1-\xp)Q^2}
\sum_{i,j} f_i \left(\frac{\xb}{\xp},Q^2\right)\, L_{\mu\nu}\,M_{ij}^{\mu\nu}
\,
D_j^h\left(z_h + \frac{\xp P_T^2}{z_h(1-\xp)Q^2},Q^2\right)\,
\label{LOxs}
\ee
with \cite{chay}
\bea
L_{\mu\nu} M_{qq}^{\mu\nu} &=&\frac{64\pi}{3}Q^2
\frac{(l\cdot k)^2+(l^\prime \cdot k^\prime)^2 +
      (l^\prime \cdot k)^2 + (l \cdot k^\prime)^2}
     {(k\cdot k^{\prime\prime})(k^\prime \cdot k^{\prime\prime})} \nonumber \\
&=& \frac{64\pi}{3}\, \frac{Q^2}{y^2} \, \left\{
[ 1 + (1-y)^2]\;
\left[(1-\xp) (1-\zp) + \frac{1+ (\xp\zp)^2}{(1-\xp)(1-\zp)}\right]
\,+\,8\,\xp\zp\;(1-y)
\right.\nonumber \\
&& \left.-\; 4\,\sqrt{\frac{\xp\zp\;(1-y)}{(1-\xp)(1-\zp)}}\;(2-y)\;[\xp\zp
+ (1-\xp)(1-\zp)]\cos\phi_h \right.\nonumber \\
  &&  \left. + \;4\,\xp\zp\,(1-y)\,\cos 2\phi_h \frac{}{}\right\} \;,
\label{LOxs1}
\eea
\bea
L_{\mu\nu} M_{qg}^{\mu\nu}&=&\frac{64\pi}{3}Q^2
\frac{(l\cdot k)^2+(l^\prime \cdot k^{\prime\prime})^2 +
      (l^\prime \cdot k)^2 + (l \cdot k^{\prime\prime})^2}
     {(k\cdot k^\prime)(k^\prime k^{\prime\prime})} \nonumber \\
&=& \frac{64 \pi}{3}\,\frac{Q^2}{y^2}\,\left\{
[ 1 + (1-y)^2]\;
\left[(1-\xp)\zp + \frac{1+{\xp}^2(1-\zp)^2}{(1-\xp)\zp}\right]
\,+\,8\,\xp(1-y)(1-\zp)
\right.\nonumber \\
&& \left.+\; 4\,\sqrt{\frac{\xp(1-y)(1-\zp)}{(1-\xp)\zp}}\;(2-y)\;[\xp(1-\zp)
+ (1-\xp)\zp]\cos\phi_h \right.\nonumber \\
  &&  \left. + \;4\,\xp(1-y)(1-\zp)\,\cos 2\phi_h \frac{}{}\right\} \;,
\label{LOxs2}
\eea
\bea
L_{\mu\nu} M_{gq}^{\mu\nu}&=&\frac{64\pi}{3}Q^2
\frac{(l\cdot k^{\prime\prime})^2+(l^\prime \cdot k^\prime)^2 +
      (l^\prime \cdot k^{\prime\prime})^2 + (l \cdot k^\prime)^2}
     {(k\cdot k^\prime)( k\cdot k^{\prime\prime})} \nonumber \\
&=&  8\pi\alpha_s\;\frac{Q^2}{y^2}\left\{
[ 1 + (1-y)^2]\;[{\xp}^2+(1-\xp)^2]\;
\frac{{\zp}^2+(1-\zp)^2}{\zp(1-\zp)}
\,+\,16\,\xp (1-\xp) (1-y)
\right.\nonumber \\
&& \left.-\; 4\,\sqrt{\frac{\xp(1-\xp)(1-y)}{\zp(1-\zp)}}\;
(2-y)\;(1-2\xp)(1-2\zp)\,\cos\phi_h \right.\nonumber \\
&&  \left. + \;8\,\xp(1-\xp)(1-y)\,\cos 2\phi_h \frac{}{}\right\} \;,
\label{LOxs3}
\eea
where we have explicitely written the scalar products in terms of $\xp$, $y$,
$\zp$ and $\cos\phi_h$. Notice the appearance of the $\cos\phi_h$ and
$\cos2\phi_h$ terms: $\phi_h$ is the azimuthal angle of the fragmenting
partons, which, in a collinear configuration, concides with the azimuthal
angle of the detected final hadron. The above expressions agree with
results previously obtained in the literature \cite{mendez-kroll}.

Large values of $P_T$ cannot be generated by the modest amount of intrinsic
motion \cite{sidis1}; we expect that Eq.(\ref{LOxs}) will dominantly describe
the cross sections for the lepto-production of hadrons with $P_T$ values
above $1$ (GeV/c).

Let us finally briefly consider the contributions of order $\alpha_s^2$ (NLO),
$d\sigma_2$ in Eq. (\ref{pert-exp}). These, for the production of large $P_T$
hadrons, have been recently computed \cite{maniatis,sassot}, resulting in large
corrections to the $\mathcal{O}(\alpha_s)$ (LO) results, and leading to a good
agreement with experimental data. It would be unnecessarily complicated, for
our purposes, to take exactly into account these contributions, as we have done
for the LO ones, via Eqs. (\ref{LOxs})--(\ref{LOxs3}).  The most simple way of
inserting the NLO results in our study is via the $K$ factor, defined as the
NLO to LO ratio of the SIDIS cross sections. A close examination of the $K$
factor shows a clear dependence on $P_T$ and the other kinematical variables:
for example, it may be larger than $10$ at low $P_T$ and $Q^2$, while
approaching unity at larger values (see, for example, Fig.~3 of
Ref.~\cite{maniatis}). However, we shall need to use the $K$ factor only in
limited ranges of $P_T$ (up to 3 GeV/$c$ at most) and $Q^2$, depending on the
sets of experimental data we consider: in these limited kinematical regions we
find that a satisfactory description of the SIDIS experimental data can be
achieved by using constant values of $K$, which will be indicated in each case.

Therefore, we will effectively include $\mathcal{O}(\alpha _s ^2)$
contributions in our computations by writing Eq.~(\ref{pert-exp}) as
\be
d\sigma = \alpha_s ^0 d\sigma_0 +  \
          \alpha_s ^1 \, K \,d\sigma_1 + ...\,,
\label{pert-exp-K}
\ee
where $d\sigma_0$ and $d\sigma_1$ are calculated according to
Eqs.~(\ref{sidis-Xsec-final}) and (\ref{LOxs}) respectively.
The first term dominates at $P_T \lsim 1$ GeV/$c$, and the second one at
$P_T \gsim 1$ GeV/$c$.

\section{\label{results} Numerical Results}

In Ref.~\cite{sidis1} several sets of experimental data, showing the explicit
dependence of the SIDIS unpolarized cross sections on the azimuthal angle $\phi
_h$ and on $P_T$, were considered. A comprehensive fit, based on Eq.
(\ref{sidis-Xsec-final}) -- or its simplified version valid up to
$\mathcal{O}(\kt/Q)$, Eq. (\ref{cahn-anal-app}) -- was performed, in order to
determine the values of $\langle\kt^2\rangle$ and $\langle\ptq^2\rangle$,
obtaining
\be
\nonumber
\langle\kt^2\rangle   = 0.25  \;({\rm GeV})^2 \quad\quad
\langle\ptq^2\rangle  = 0.20 \;({\rm GeV})^2 \>. \label{oldpar}
\ee
These values were assumed to be constant and flavour independent.
%
\begin{figure}
\begin{center}
\includegraphics[width=0.5\textwidth,bb= 10 140 540 660,angle=-90]
{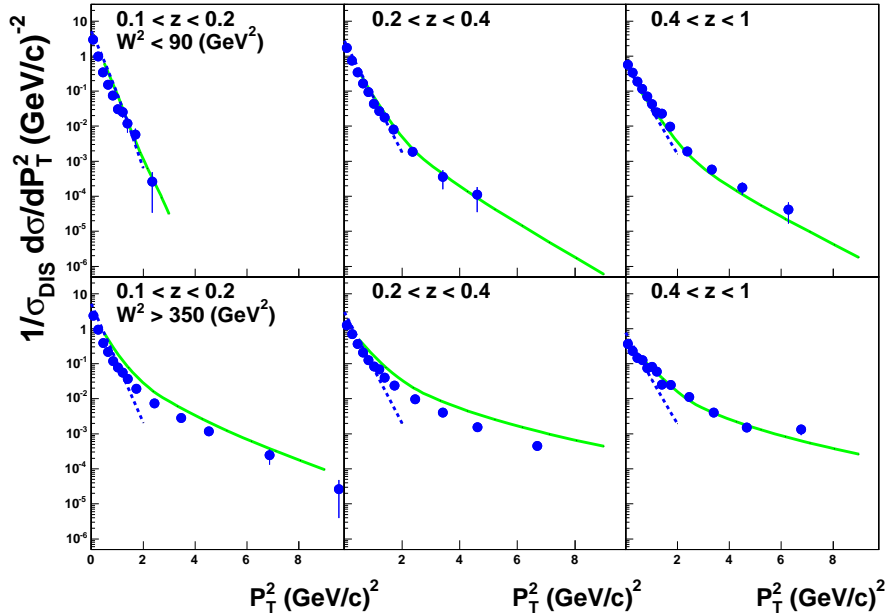}
\end{center}
\caption{\label{ptdist} The normalized cross section $d\sigma/dP_T^2$: the
dashed line reproduces the $\mathcal{O}(\alpha_s^0)$ contributions, computed by
taking into account the partonic transverse intrinsic motion at all orders in
$(k_\perp/Q)$, Eq. (\ref{sidis-Xsec-final}). The solid line corresponds to
collinear and pQCD contributions, computed at LO, with a $K$ factor ($K=6$) to
account for NLO effects, Eqs. (\ref{LOxs})-(\ref{pert-exp-K}). The data are
from EMC collaboration measurements \cite{EMCpt}. $\langle k_\perp\rangle$ and
$\langle p_\perp\rangle$ are fixed as in Eq.~(\ref{parameters}).}
\end{figure}

The study of Ref. \cite{sidis1} also showed clearly that Eq.
(\ref{sidis-Xsec-final}) -- zeroth order pQCD with TMD distribution and
fragmentation function -- works very well in the $P_T \simeq \Lambda_{\rm QCD}
\simeq \kt$ region, but fails at larger $P_T$ values, where higher order pQCD
contributions, with collinear partonic configurations, are expected to take
over and explain the data. The transition point is around $P_T \simeq 1$
GeV/$c$.

We have redone the analysis of Ref. \cite{sidis1} taking into account,
in the appropriate kinematical regions, also the pQCD contributions.
It turns out that, while a complete description of the data in the full
$P_T$ range is possible, a little variation is required to the values
given in Eq. (\ref{oldpar}). Actually, the resulting change is included within
a 20\% variation of the parameters, already considered in Ref. \cite{sidis1}.

\begin{figure}[t]
\begin{center}
\includegraphics[width=0.3\textwidth,bb= 10 140 540 660,angle=-90]
{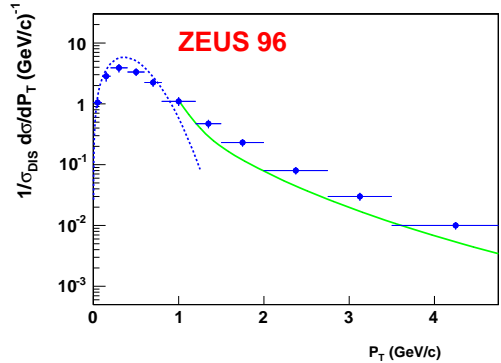}
\end{center}
\caption{\label{ptdist-zeus} The normalized cross section $d\sigma/dP_T$: the
dashed line reproduces the $\mathcal{O}(\alpha_s^0)$ contribution, computed by
taking into account the partonic transverse intrinsic motion at all orders in
the $(k_\perp/Q)$ expansion, Eq. (\ref{sidis-Xsec-final}); the solid line
corresponds to the SIDIS cross section as given by LO contributions and a $K$
factor ($K=1.5$) to account for NLO effects, Eqs.
(\ref{LOxs})--(\ref{pert-exp-K}). The data are from ZEUS collaboration
measurements \cite{Derrick96}. $\langle k_\perp\rangle$ and $\langle
p_\perp\rangle$ are fixed as in Eq.~(\ref{parameters}).}
\end{figure}

\begin{figure}[t]
\includegraphics[width=0.4\textwidth,bb= 0 170 540 660, angle=-90]
{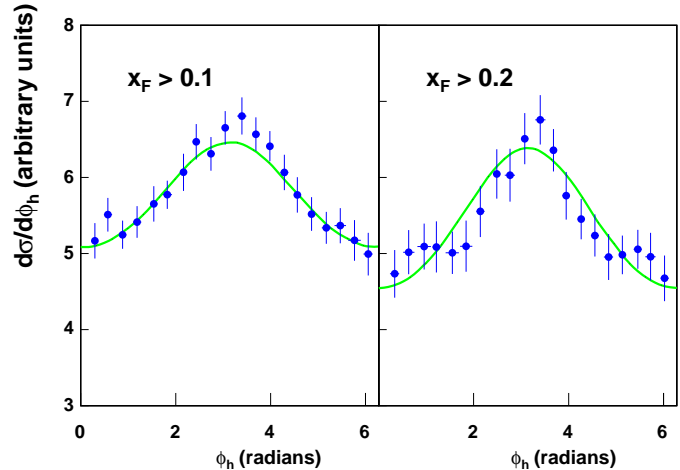}
\caption{\label{ds/dphi} The cross section $d\sigma/d\phi_h$: the solid line is
obtained by including all orders in $(k_\perp/Q)$, the LO corrections and a
$K=6$ factor to account for NLO effects. The data are from EMC measurements
\cite{EMC2}. $\langle k_\perp\rangle$ and $\langle p_\perp\rangle$ are fixed as
in Eq.~(\ref{parameters}). }
\end{figure}

\begin{figure}
\begin{center}
\includegraphics[width=0.3\textwidth,bb= 10 140 540 660,angle=-90]
{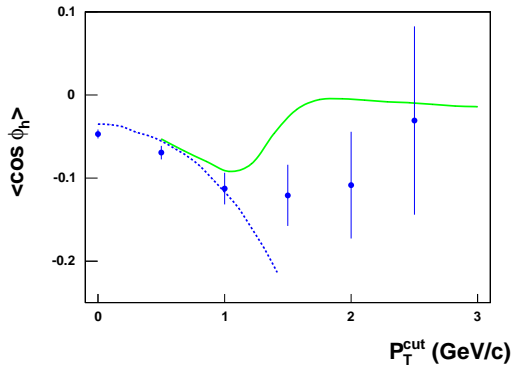}
\end{center}
\caption{\label{cosphi} $\langle \cos\phi_h\rangle$ as a function of
$P_T^{cut}$: the dashed line reproduces the $\mathcal{O}(\alpha_s^0)$
contribution, computed by taking into account the partonic transverse intrinsic
motion at all orders in $(k_\perp/Q)$; the solid line corresponds to the SIDIS
cross section as obtained by including LO corrections and a $K=6$ factor to
account for NLO effects. The data are from the E665 collaboration \cite{E665}.}
\end{figure}

\begin{figure}
\begin{center}
\includegraphics[width=0.3\textwidth,bb= 10 140 540 660,angle=-90]
{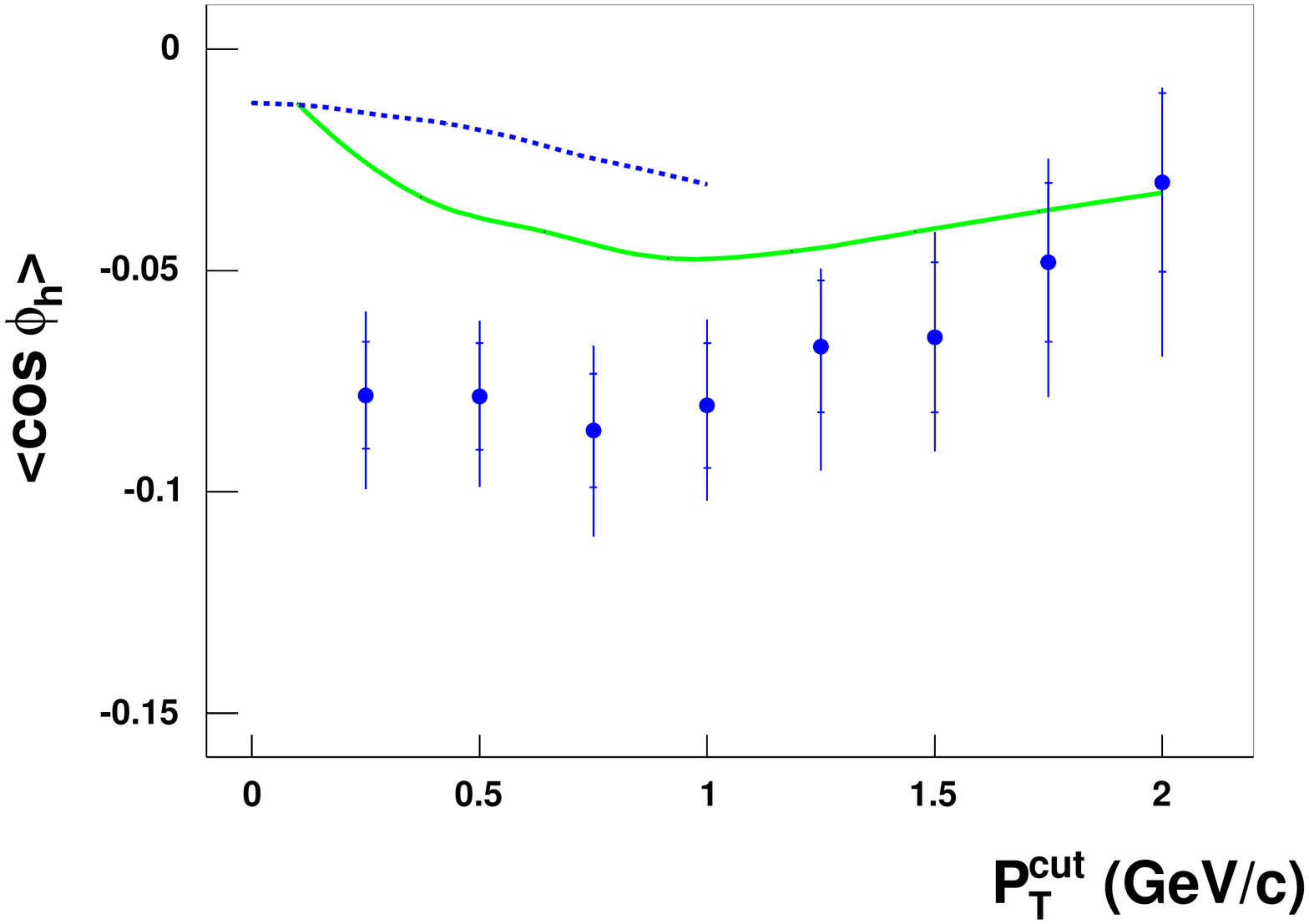}
\end{center}
\caption{\label{cosphi-zeus} $\langle \cos\phi_h\rangle$ as a function of
$P_T^{cut}$: the dashed line reproduces the $\mathcal{O}(\alpha_s^0)$
contribution, computed by taking into account the partonic transverse intrinsic
motion at all orders in $(k_\perp/Q)$; the solid line corresponds to the SIDIS
cross section as obtained by including LO corrections and a  $K = 1.5$  factor
to account for NLO effects. The data are from the ZEUS collaboration
\cite{Breitweg}.}
\end{figure}

Figs.~\ref{ptdist}-\ref{cosphi-zeus} show our results obtained by adding,
according to Eq. (\ref{pert-exp-K}), the contributions of Eq.
(\ref{sidis-Xsec-final}) to the contributions (computed above $P_T = 1$
GeV/$c$) of Eq. (\ref{LOxs}). We have used
\be
\nonumber
\langle\kt^2\rangle   = 0.28  \;({\rm GeV})^2 \quad\quad
\langle\ptq^2\rangle  = 0.25 \;({\rm GeV})^2 \>,
\label{parameters}
\ee
again constant and flavour independent. The $K$ factor was fixed to be a
constant, with different values according to the different $P_T$ and $Q^2$
ranges involved. Here and throughout the paper we have adopted the MRST01 NLO
\cite{MRST01} set of distribution functions and the fragmentation functions
by Kretzer \cite{Kretzer} at NLO.

Let us comment in greater detail on each single plot. In Fig.~\ref{ptdist} we
compare our results to the EMC measurements of the SIDIS $P_T^2$ distributions
(normalized to the integrated DIS cross section) \cite{EMCpt}, defined as
\be
\frac{1}{\sigma_{_{\rm DIS}}}\frac{d\sigma}{dP_T^2} = \frac{1}{2\sigma_{_{\rm DIS}}}\,
\int \! d\phi_h \, d\xb \, dy \, dz_h
\frac{d^5\sigma^{\ell p \to \ell h X }}{d\xb \, dy \, dz_h \, d^2 \bfP _T},
\label{dsigma/dPT2}
\ee
where the integration covers the $\xb$, $y$, $z_h$ and $P_T$ regions consistent
with the experimental cuts:
\be \nonumber Q^2 > 5 \; ({\rm GeV}/c)^2 \quad\quad E_h > 5 \; ({\rm GeV})
\quad\quad 0.1 < z_h < 1 \quad\quad 0.2 < y < 0.8\,. \ee
The dashed lines reproduce the $\mathcal{O}(\alpha_s^0)$ contribution, computed
by taking into account the partonic transverse intrinsic motion at all orders
in $(k_\perp/Q)$, whereas the solid lines correspond to the SIDIS cross section
as obtained by including LO corrections and the $K$ factor ($K=6$) to account
for NLO effects. The two contributions together give a very good complementary
description of the data over the full $P_T$ domain. Unavoidably, there is a
slight mismatch at the transition point, $P_T = 1$ GeV/$c$, where both
contributions somewhat describe the same physics, and some kind of average
should be performed to avoid double counting. The value $K=6$ is the simplest,
although rough, approximation, in the kinematical range of the data considered
here, to the computed $K$ factor (see, Fig. 3 of Ref. \cite{maniatis}).
$\sigma_{_{\rm DIS}}$ is evaluated starting from Eq. (17) of Ref.
\cite{sidis1}; pQCD corrections for this integrated quantity are negligible.

A similar very good agreement, shown in Fig.~\ref{ptdist-zeus}, is obtained
when comparing our computations with the experimental measurements of the
ZEUS collaboration at DESY \cite{Derrick96}. Here the SIDIS differential
cross section,
\be
\frac{1}{\sigma_{_{\rm DIS}}}\frac{d\sigma}{dP_T} = \frac{1}{\sigma_{_{\rm DIS}}}\,
\int \! d\phi_h \; d\xb dQ^2 dz_h P_T
\frac{d^5\sigma^{\ell p \to \ell h X }}{d\xb \, dQ^2 \, dz_h \, d^2 \bfP _T}
\>, \label{dsigma/dPT}
\ee
is obtained by performing the integrations according to the following
experimental conditions
\be 10<Q^2<160\;({\rm GeV}/c)^2 \quad\quad 75<W<175 \;({\rm GeV}) \>, \ee
and the $K$ factor is taken to be 1.5 (as shown in Fig. 8 of Ref.
\cite{maniatis}).

In Fig.~\ref{ds/dphi} we compare our results to the EMC measurements of
the (not normalized) $\phi_h$ distributions \cite{EMC2}, proportional to
\be
\frac{d\sigma ^{\ell p \to \ell h X}}{d\phi_h}  =
 \int \! d\xb \, dy \, dz_h \,  P_T \, d P _T \;
\frac{d^5\sigma^{\ell p \to \ell h X }}{d\xb \, dy \, dz_h \, d^2 \bfP _T}\,.
\label{eq:ds/dphi}
\ee
The integration regions are fixed by the conditions
\be \nonumber x_{F} > 0.1 \quad\quad P_T > 0.2 \; ({\rm GeV}/c) \quad\quad 0.2
< y < 0.8 \quad\quad Q^2>\; 4({\rm GeV}/c)^2\,, \ee
where $x_{F} = {2 P_L}/{W}\,$ and $P_L$ is the longitudinal momentum of the
produced hadron relative to the virtual photon.
%
%
%

This is a quantity integrated over $P_T \ge 0.2$ GeV/$c$; we have used only
$d\sigma_0$ up to $P_T = 1$ GeV/$c$ and added the $K \,d\sigma_1$ contributions
(with $K = 6$) above that. We notice, however, that the dominant contributions
come from very low $P_T$'s, while the pQCD contribution is almost negligible.

Figs.~\ref{cosphi} and \ref{cosphi-zeus} show our predictions for the average
value of $\cos\phi_h$ compared to the experimental data from the FNAL E665
collaboration \cite{E665} ($\mu p$ and $\mu d$ interactions at $490$ GeV) and
from the ZEUS collaboration \cite{Breitweg} (positron-proton collisions at
$300$ GeV) respectively. Here $\langle\cos\phi_h\rangle$ is defined as
\be \langle\cos\phi_h\rangle  = \frac{\int d\xb dQ^2 dz_h d^2 \bfP _T \;
\cos\phi_h \; d^5\sigma} {\int d\xb dQ^2 dz_h d^2 \bfP _T  \; d^5\sigma}\,,
\label{eq:cosphi} \ee
where $d^5\sigma$ denotes the fully differential cross section
\be d^5\sigma \equiv \frac{d^5\sigma ^{\ell p \to \ell h X }} {d\xb \, dQ^2 \,
dz_h \, d^2 \bfP _T} \, \cdot \ee
For the FNAL E665 data sample the integral over $P_T$ runs from $P_T^{cut}$ to
$P_T^{max}\sim 10$~GeV/$c$ and the range of the other variables is fixed by the
following experimental cuts:
\bea && Q^2 > 3 \; {\rm (GeV)^2}\;,\;\;\; 300 < W^2 < 900 \; {\rm (GeV)^2}\;,
\nonumber \\
&& 60 < \nu < 500 \; {\rm (GeV)}\;,\;\;\; E_h > 8 \; {\rm (GeV)} \;,\;\;\; 0.1
< y < 0.85\,, \eea whereas for the ZEUS data sample the integral over $P_T$
runs from $P_T^{cut}$ to $P_T^{max}\sim 10$~GeV/$c$, within the ranges
\bea \nonumber
&& 0.01 < \xb < 0.1 \;,\;\;\;0.2 < y < 0.8 \;,\nonumber\\
&& 0.2 < z_h < 1.0\;,\;\;\; 180 < Q^2 < 7220 \; {\rm (GeV)^2}\;. \eea
As in the previous case, we have added perturbative corrections only from $P_T
= 1$ (GeV/$c$), leaving $d\sigma_0$ to be the only contributing term for values
of $P_T$ below $1$ (GeV/$c$).

The results we obtain are in good qualitative agreement with the FNAL E665
experimental data. As expected, they show that the pQCD contributions are very
small at low $P_T^{cut}$ values, but quickly increase as $P_T^{cut}$ raises,
significantly correcting the fast fall of the $\mathcal{O}(\alpha_s^0)$ term,
as shown in Fig.~\ref{cosphi}.

Instead, our results disagree with the ZEUS data, especially in the lower range
of $P_T^{\rm cut}$. This is surprising and would deserve further experimental
studies. The $\cos\phi_h$ modulation, at small $P_T$ values, is a kinematical
higher-twist effect, and decreases like $P_T/Q$ for growing values of $Q$, as
shown in Eq.~(\ref{cahn-anal-app}). Therefore we expect, and indeed we find,
$\langle \cos\phi_h\rangle$ to be much smaller for ZEUS data, which correspond
to huge values of $Q^2$, than for E665 results, which correspond to much lower
$Q^2$ values.

\begin{figure}
\begin{center}
\includegraphics[width=0.3\textwidth,bb= 10 140 540 660,angle=-90]
{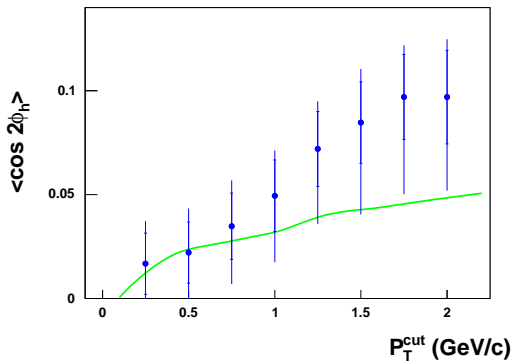}
\end{center}
\caption{\label{cos2phi-zeus} $\langle \cos(2\phi_h)\rangle$ as a function of
$P_T^{cut}$ as obtained by including LO corrections and a $K=1.5$ factor to
account for NLO effects. The data are from the ZEUS collaboration
\cite{Breitweg}.}
\end{figure}

We have also computed, with the same procedure, $\langle \cos(2\phi_h)\rangle$.
We have seen that such a dependence can arise both, at $\mathcal{O}(P_T/Q)^2$,
from intrinsic motion, and, at $\mathcal{O}(\alpha_s)$, from pQCD corrections.
However, there is another non perturbative, leading-twist, small $P_T$ source
of the $\cos(2\phi_h)$ dependence, related to the combined action of the
Boer-Mulders \cite{bm} and Collins \cite{col} effects; this has been recently
studied in Ref \cite{bar}, where both the $1/Q^2$ kinematical contribution and
the Boer-Mulders $\otimes$ Collins one were studied and found to be of
comparable size. Therefore, our results, which ignore the Boer-Mulders
$\otimes$ Collins contribution, can be considered reliable only in the large
$P_T$ region; indeed, in this region, we find agreement with the ZEUS
experimental data of Ref.~\cite{Breitweg}, as shown in Fig.~\ref{cos2phi-zeus}.
%
%
%
%
%
%
%

\newpage

\section{\label{predictions} Predictions for forthcoming measurements}

New data are expected from ongoing measurements or data analysis at HERMES,
COMPASS and JLab. They concern dominantly the small $P_T$ (and, hopefully,
large enough $Q^2$) region, where we have seen that the simple partonic
approach, with unintegrated distribution and fragmentation functions, can give
a very satisfactory description of the available data. We can easily give
detailed predictions which can soon be tested, allowing a further check on the
role of intrinsic motions in affecting physical observables. We consider the
SIDIS cross sections and the average value of $\cos\phi_h$. Also $\langle
\cos2\phi_h\rangle$, keeping in mind the comments at the end of the previous
Section, is computed.

\begin{figure}[t]
\begin{center}
\includegraphics[width=0.85\textwidth]{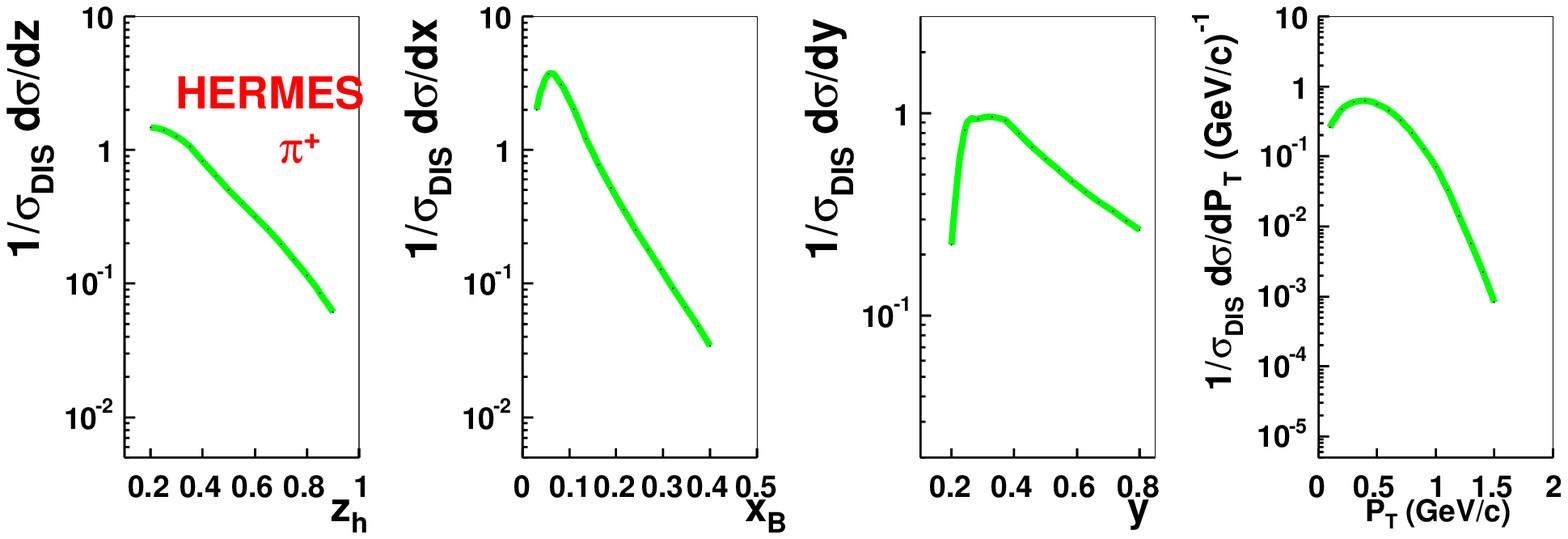}
\end{center}
\caption{\label{pt-hermes} Predictions for the normalized SIDIS cross section
corresponding to the production of $\pi^+$
as it will be measured by the HERMES collaboration in the forthcoming future.
The solid lines correspond
to the SIDIS cross section as obtained by including all orders in the
$(k_\perp/Q)$ expansion. Notice that QCD corrections have no influence in
this range of low $P_T$'s.}
\end{figure}

\begin{figure}[t]
\begin{center}
\vspace*{1.2cm}
\includegraphics[width=0.8\textwidth]
{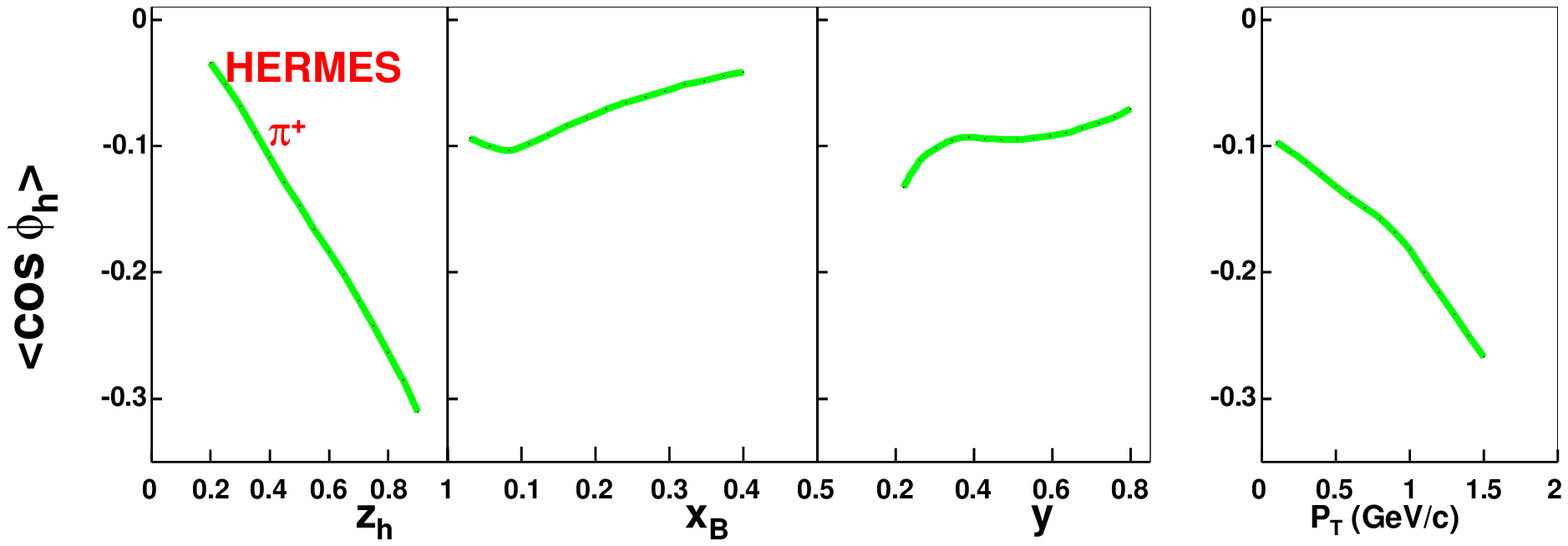}
\end{center}
\caption{\label{cosphi-hermes} Predictions for $\langle \cos\phi_h\rangle$
corresponding to the production of $\pi^+$
as it will be measured by the HERMES collaboration in the forthcoming future.
 The solid lines correspond
to $\langle \cos\phi_h\rangle$ we find by including all orders in the
$(k_\perp/Q)$ expansion. Notice that QCD corrections have no influence in
this range of low $P_T$'s.}
\end{figure}

\begin{figure}[t]
\begin{center}
\includegraphics[width=0.3\textwidth,bb= 0 200 540 660,angle=-90]
{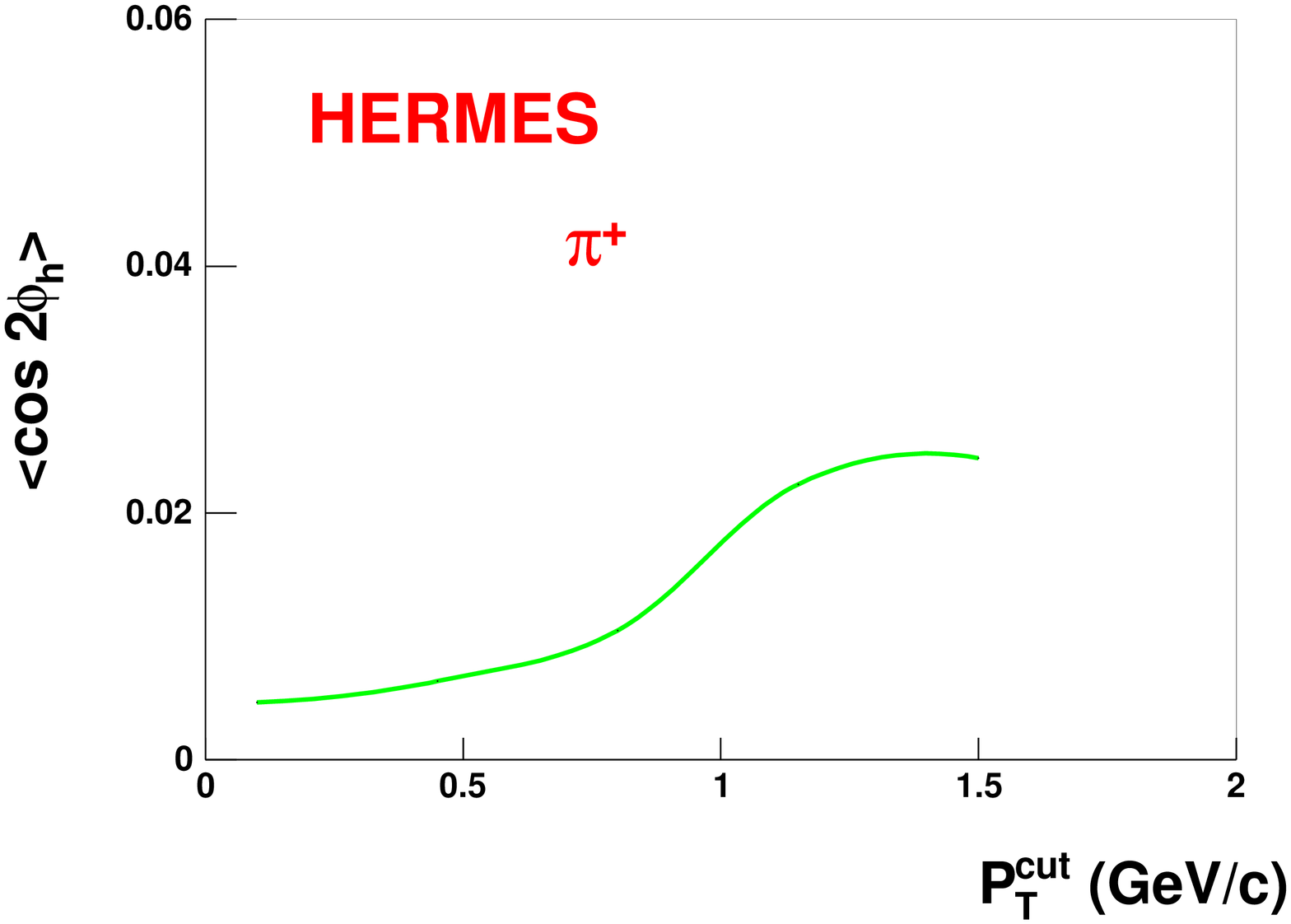}
\end{center}
\caption{\label{cos2phi-hermes} Predictions for $\langle \cos 2\phi_h\rangle$
corresponding to the production of $\pi^+$
as it will be measured by the HERMES collaboration in the forthcoming future.
 The solid lines correspond
to $\langle \cos2\phi_h\rangle$ we find by including all orders in the
$(k_\perp/Q)$ expansion. Notice that QCD corrections have no influence in
this range of low $P_T$'s.}
\end{figure}

In Fig.~\ref{pt-hermes}, we plot the SIDIS cross section, for $\pi^+$
production at HERMES, as function of one variable at a time, either $z_h$,
$\xb$, $y$ or $P_T$; the integration over the unobserved variables has been
performed consistently with the setup of the HERMES experiment, which studies
the scattering of positrons at $p_{lab} = 27.57$ GeV/$c$ against a fixed
hydrogen gas target:
\bea \nonumber && Q^2 > 1 \; ({\rm GeV}/c)^2 \quad\quad W^2 > 10 \; {\rm GeV}^2
\quad\quad P_T > 0.05 \; {\rm GeV}/c \\
\label{hermutcuts} && 0.023 < \xb < 0.4 \quad\quad 0.2 < z_h < 0.7 \quad\quad
0.1 < y < 0.85 \quad\quad 2 < E_h < 15 \> {\rm GeV} \>. \eea
In these kinematical regions the cross section is heavily dominated by the
$d\sigma_0$ term of Eq. (\ref{pert-exp}), computed according to
Eq.~(\ref{sidis-Xsec-final}); $\sigma_{DIS}$ is computed according to Eq. (17)
of Ref. \cite{sidis1}. We also evaluate the average value of $\cos\phi_h$ in
the same kinematical region, shown in Fig. \ref{cosphi-hermes}, and of
$\cos2\phi_h$, shown in Fig.~\ref{cos2phi-hermes}. The latter, however, can
only be taken as a partial (higher-twist) contribution to the real $\langle
\cos2\phi_h \rangle$, as explained at the end of the last section.

We notice that very similar predictions, for the cross section, $\langle
\cos\phi_h \rangle$ and $\langle \cos2\phi_h \rangle$, are obtained for $\pi^0$
and $\pi^-$ production, which we do not show.

The COMPASS experiment at CERN collects data in $\mu d \to \mu h^\pm X$
processes at $p_{lab}=$ 160 GeV/$c$, covering the following kinematical
regions:
\bea
\nonumber
&& Q^2 > 1 \;  ({\rm GeV}/c)^2 \quad W^2 > 25 \; {\rm GeV}^2 \quad
P_T > 0.1 \; {\rm GeV}/c \\ &&
E_h < 15 \;{\rm GeV} \quad\quad
0.2 < z_h < 1 \quad\quad  0.1 < y < 0.9 \> .
\label{compass-cut}
\eea

\begin{figure}
\begin{center}
\includegraphics[width=0.85\textwidth]
{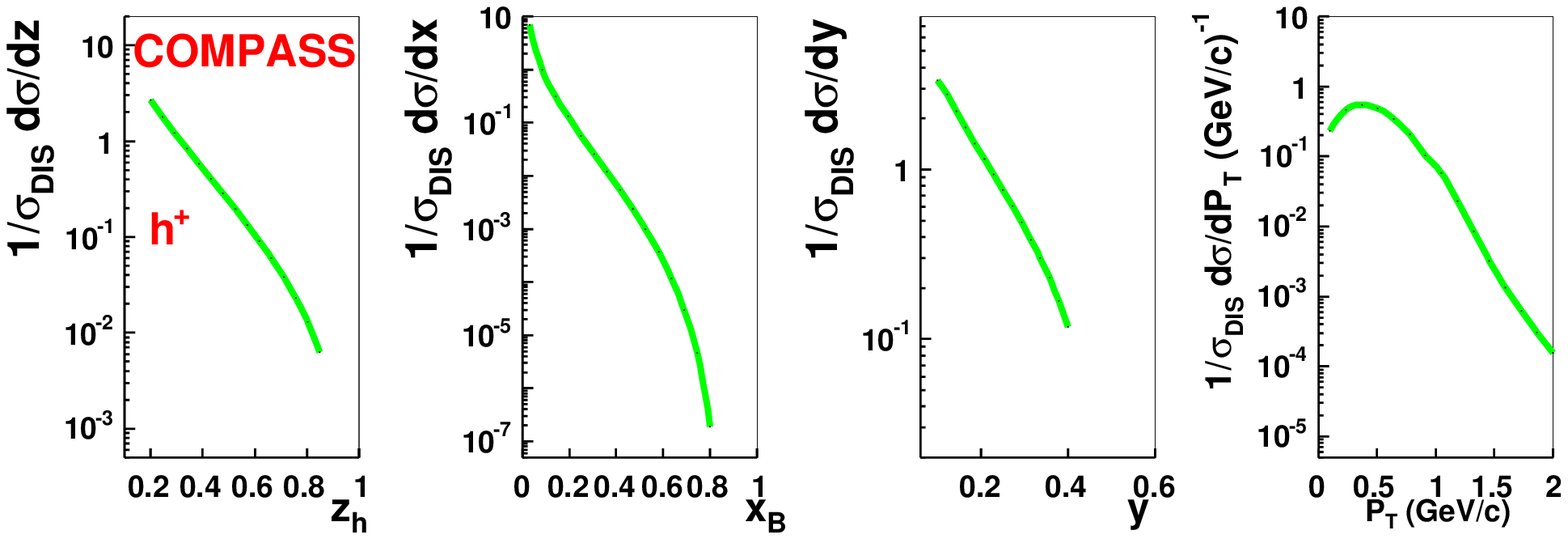}
\end{center}
\caption{\label{pt-compass} Predictions for the normalized SIDIS cross section
corresponding to the production of positively charged hadrons
as it will be measured by the COMPASS collaboration in the forthcoming future.
The solid lines correspond
to the SIDIS cross section as obtained by including all orders in the
$(k_\perp/Q)$ expansion. Notice that QCD corrections have no influence in
this range of low $P_T$'s.}
\end{figure}

\begin{figure}
\begin{center}
\includegraphics[width=0.8\textwidth]
{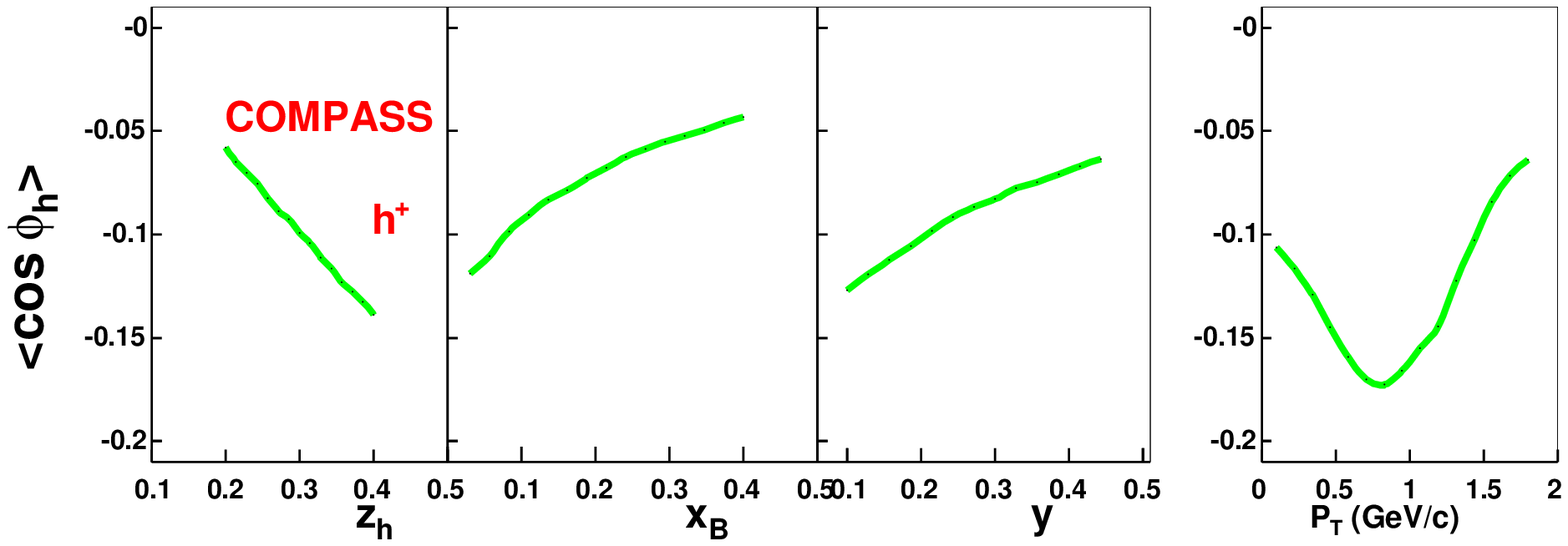}
\end{center}
\caption{\label{cosphi-compass}  Predictions for $\langle \cos\phi_h\rangle$
corresponding to the production of positively charged hadrons
as it will be measured by the COMPASS collaboration in the forthcoming future.
The solid lines correspond
to $\langle \cos\phi_h\rangle$ we find by including all orders in the
$(k_\perp/Q)$ expansion. Notice that QCD corrections have no influence in
this range of low $P_T$'s.}
\end{figure}

\begin{figure}
\begin{center}
\includegraphics[width=0.3\textwidth,bb= 0 200 540 660,angle=-90]
{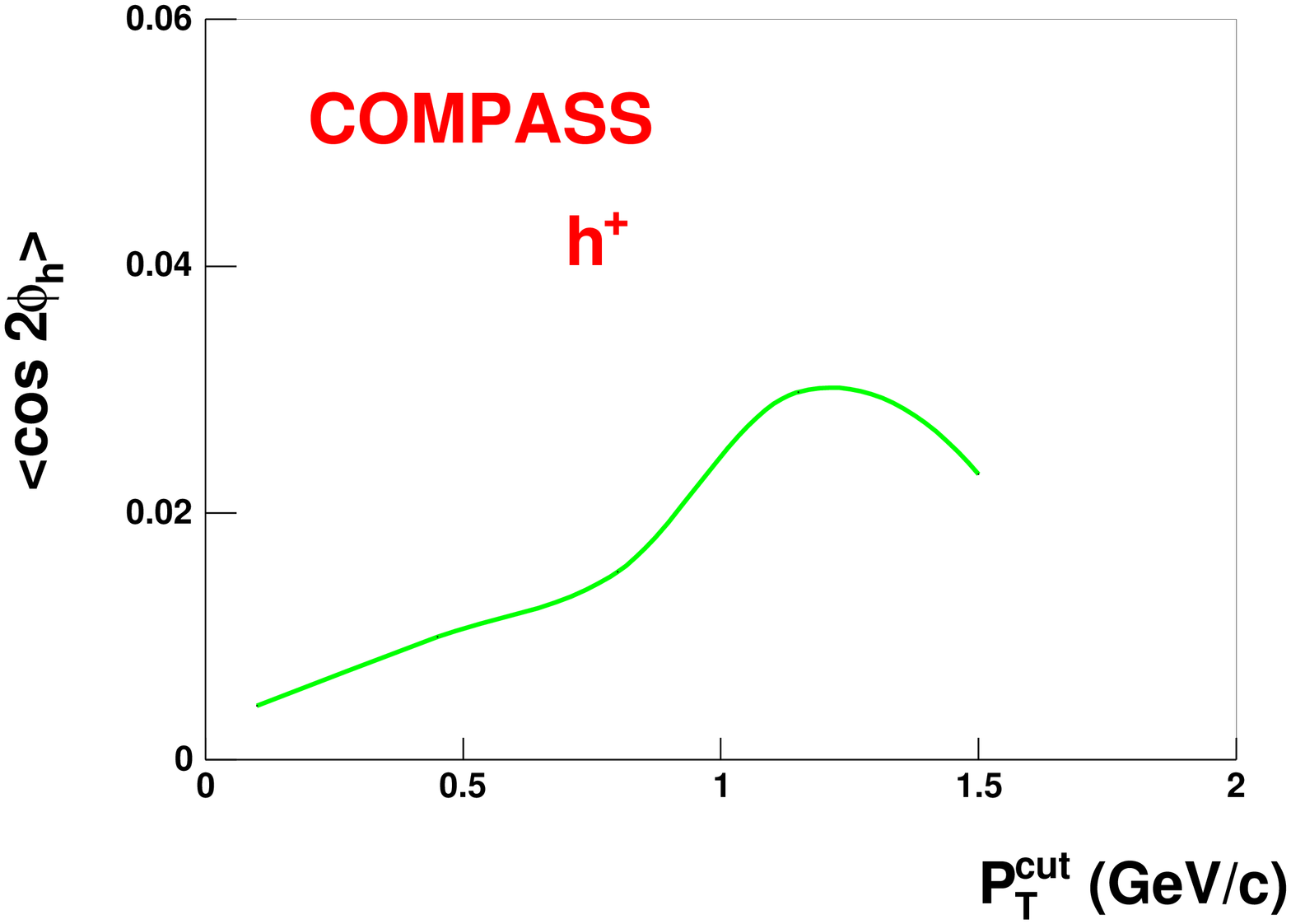}
\end{center}
\caption{\label{cos2phi-compass}  Predictions for $\langle \cos2\phi_h\rangle$
corresponding to the production of positively charged hadrons
as it will be measured by the COMPASS collaboration in the forthcoming future.
The solid lines correspond
to $\langle \cos2\phi_h\rangle$ we find by including all orders in the
$(k_\perp/Q)$ expansion. Notice that QCD corrections have no influence in
this range of low $P_T$'s.}
\end{figure}

Fig.~\ref{pt-compass} shows our corresponding predictions for the SIDIS cross
section -- for the production of positively charged hadrons -- as a function of
the kinematical variables $z_h$, $\xb$, $y$ and $P_T$, as obtained from
Eq.~(\ref{sidis-Xsec-final}). Notice that we neglect nuclear corrections and
use the isospin symmetry in order to obtain the parton distribution functions
of the deuterium. Similarly to the HERMES case, the perturbative QCD
corrections are negligible. Predictions for $\langle \cos\phi_h\rangle$ are
presented in Fig.~\ref{cosphi-compass} and the higher-twist contributions to
$\langle \cos 2\phi_h\rangle$ in Fig.~\ref{cos2phi-compass}. Similar results
hold for negatively charged hadrons.

\begin{figure}
\begin{center}
\includegraphics[width=0.85\textwidth]{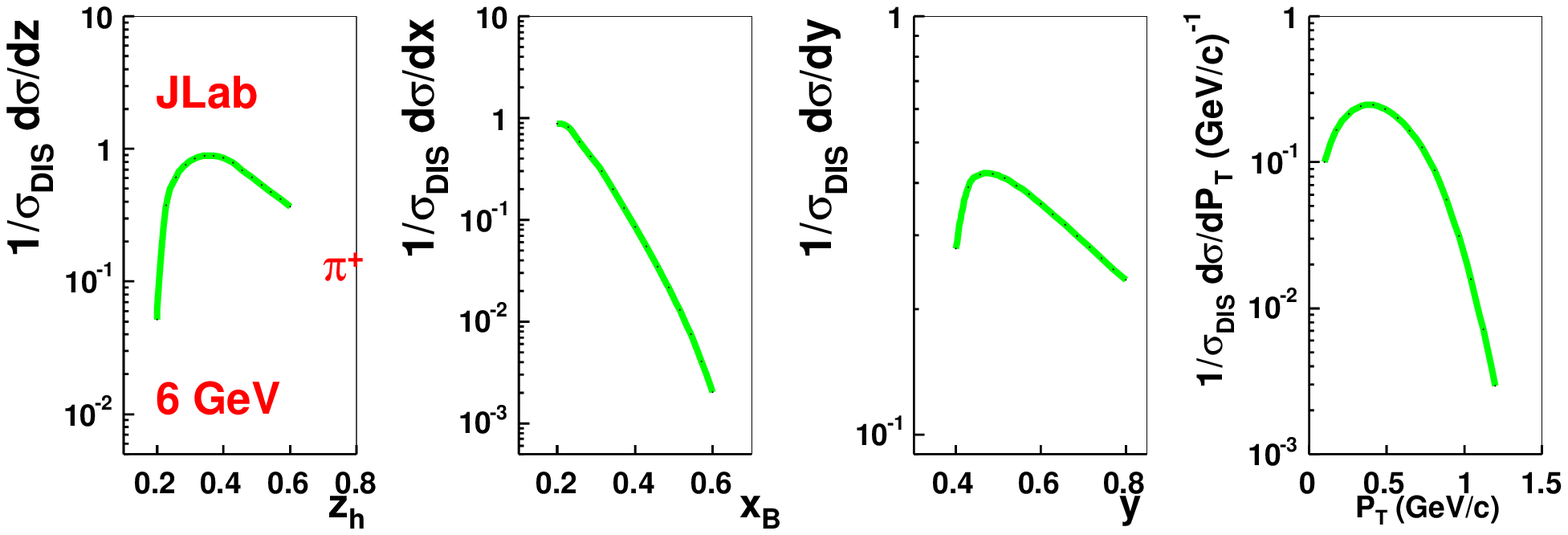}
\end{center}
\caption{\label{pt-jlab6} Predictions for the normalized SIDIS cross section
corresponding to the production of $\pi^+$
as it will be measured by the JLab collaboration in the forthcoming future.
The solid lines correspond
to the SIDIS cross section as obtained by including all orders in the
$(k_\perp/Q)$ expansion. Notice that QCD corrections have no influence in
this range of low $P_T$'s.}
\end{figure}

\begin{figure}
\begin{center}
\includegraphics[width=0.8\textwidth]
{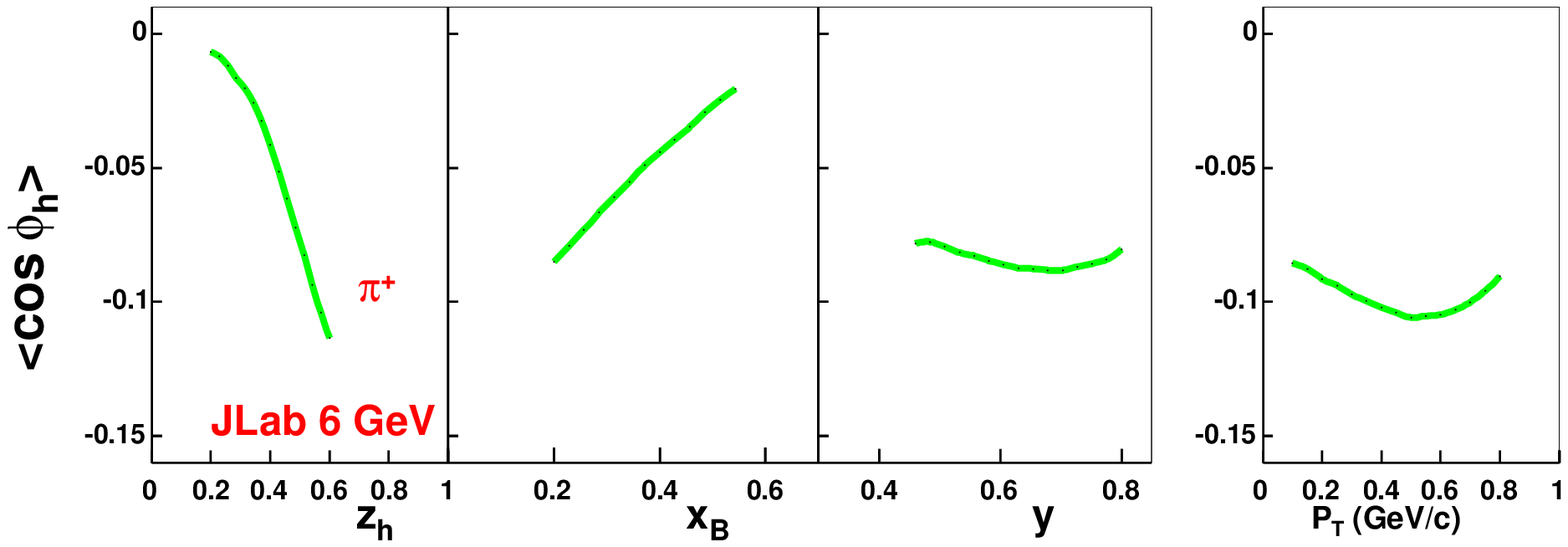}
\end{center}
\caption{\label{cosphi-jlab6} Predictions for $\langle \cos\phi_h\rangle$
corresponding to the production of $\pi^+$
as it will be measured by the JLab collaboration in the forthcoming future.
The solid lines correspond
to $\langle \cos\phi_h\rangle$ we find by including all orders in the
$(k_\perp/Q)$ expansion. Notice that QCD corrections have no influence in
this range of low $P_T$'s.}
\end{figure}

\begin{figure}
\begin{center}
\includegraphics[width=0.3\textwidth,bb= 0 200 540 660,angle=-90]
{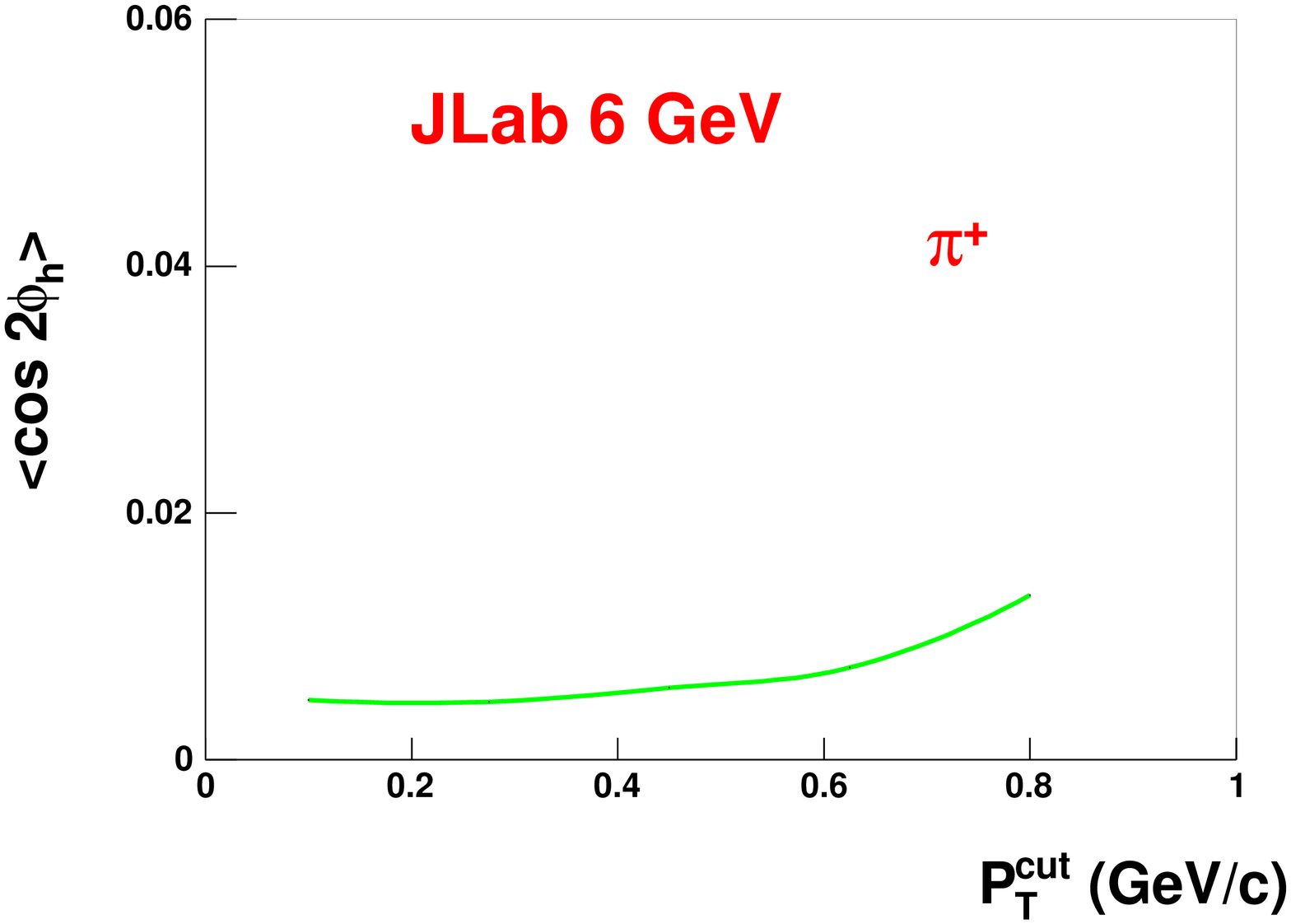}
\end{center}
\caption{\label{cos2phi-jlab6} Predictions for $\langle \cos2\phi_h\rangle$
corresponding to the production of $\pi^+$
as it will be measured by the JLab collaboration in the forthcoming future.
The solid lines correspond
to $\langle \cos2\phi_h\rangle$ we find by including all orders in the
$(k_\perp/Q)$ expansion. Notice that QCD corrections have no influence in
this range of low $P_T$'s.}
\end{figure}

Finally, JLab collects and will collect data in the collisions of $6$ and $12$
GeV electrons from a fixed He$^3$ target. In this case the relevant kinematical
regions are the following:
\bea \nonumber && Q^2 > 1 \;  ({\rm GeV}/c)^2 \quad W^2
> 4 \; {\rm GeV}^2 \quad P_T
> 0.1 \; {\rm GeV}/c \\ && 1.0 < E_h < 3.5 \;{\rm GeV} \quad\quad 0.4 < z_h <
0.7 \quad\quad  0.1 < \xb < 0.6 \quad\quad 0.4 < y < 0.85 \> . \label{jlab-cut}
\eea
Our results for the corresponding SIDIS cross sections, $\langle
\cos\phi_h\rangle$  and $\langle \cos 2\phi_h\rangle$ are shown in
Figs.~\ref{pt-jlab6},~\ref{cosphi-jlab6}~and~\ref{cos2phi-jlab6} respectively:
as for the previous experiments, also the JLab data are dominated by
$\mathcal{O}(\alpha_s^0)$ terms and are almost completely insensitive to the LO
and NLO perturbative QCD corrections. Again, we show results for $\pi^+$
production, but very similar ones hold for $\pi^-$ and $\pi^0$, which we do not
show.

All these results depend on intrinsic momenta, both $\bfk_\perp$ in the
partonic distributions and $\bfp_\perp$ in the quark fragmentation. This is
obvious for quantities like $d\sigma/dP_T$ and $\langle \cos\phi_h \rangle$
which could not even be defined, at $\mathcal{O}(\alpha_s^0)$, without
intrinsic motion (and pQCD corrections are negligible for the experiments we
consider). However, this is also true for the differential cross-sections in
$z$, $\xb$ and $y$: although they get contributions from intrinsic motion only
at $\mathcal{O}(P_T/Q)^2$, as one can explicitly see from Eq.
(\ref{cahn-anal-app}), these contributions can be sizeable in the kinematical
domains of HERMES, COMPASS and JLab.

\section{\label{conclusions} Conclusions}

We have considered the azimuthal and $P_T$ dependence of SIDIS data from low to
large $P_T$; they both cannot be explained in the simple parton model
($\mathcal{O}(\alpha_s^0)$) with collinear configurations. They can originate
from intrinsic motions and/or pQCD corrections. The outcome of our analysis
turns out to be very simple: up to $P_T \lsim 1$ GeV/$c$ the simple parton
model with unintegrated distribution and fragmentation functions explains the
data and leads to a good evaluation of $\langle k_\perp \rangle$ and $\langle
p_\perp \rangle$, while at larger $P_T$ values, $P_T \gsim 1$ GeV/$c$, the
perturbative QCD contributions originating from hard gluonic radiation
processes and elementary scattering initiated by gluons, are dominant.

Having clearly established the complementarity of the two approaches, and in
particular having gained full confidence in the domain of applicability of the
unintegrated parton model, we have given predictions for the cross sections and
for the values of $\langle \cos\phi_h\rangle$, as they will soon be measured by
HERMES, COMPASS and JLab collaborations, mainly in the low $P_T$ region. These
new data will be a very important tool to test our knowledge of the intrinsic
partonic internal motion and on the TMD quark distribution and fragmentation
functions.

\newpage

%
%
%
%
%

\newpage

%
%
%
%

\newpage

%
%

\end{document}